# Modelling the Performance of Single-Photon Counting Kinetic Inductance Detectors

**Josie Dzifa Akua Parrianen[1], Andreas Papageorgiou[1], Simon Doyle[1] and Enzo Pascale[1,2]**

[1]School of Physics and Astronomy, Cardiff University
[2]Dipartimento di Fisica, Sapienza Universita di Roma

**Abstract** Using conventional superconductor theory we discuss and validate a model that describes the energy-resolving performance of an aluminium LEKID to single-photon absorption events. While aluminium is not the optimum material for single-photon counting applications, this material is well understood and is used to understand the underlying device physics of these detectors. We also discuss data analysis techniques used to extract single-photon detections from noisy data.

## 1 Introduction

A principal goal of the next generation of space-based astronomy will be dedicated to the characterisation of extra-solar planets (exoplanets). Of the 1000s discovered only a handful have been characterised beyond their size and mass, thus there is increasing interest in new exoplanet missions aiming to carry out spectroscopy on very low intensity light and shallow transit light curves. Energy-resolving, single-photon counting detectors provide an elegant solution for carrying out spectroscopy without the need for gratings, prisms or combinations thereof. The lumped element kinetic inductance detector (LEKID) is a proven technology capable of counting and energy-resolving single-photon events at optical and near infra-red wavelengths [1, 2]. We attempt to understand the underlying physics governing single-photon detection in LEKIDs.

## 2 Single-Photon Response Model

LEKIDs are thin-film, superconducting mirco-resonators. Photon absorption leads to the breaking of Cooper pairs to create quasiparticles (un-paired

**Josie Dzifa Akua Parrianen • Andreas Papageorgiou • Simon Doyle • Enzo Pascale**

electrons in the superconductor). The quasiparticle population determines the resonance frequency of the resonator, through its surface impedance. Changes to the surface impedance are governed by kinetic inductance.

The absorption of a photon of some energy $h\nu$ creates a proportional number of excess quasiparticles. A 1st order estimate is given by

$$N_{qp,xs} = \frac{\eta h\nu}{\Delta}, \quad (1)$$

where $\eta = 0.4$ and $\Delta$ is the superconductor energy gap. Note that the typical value taken for $\eta = 0.57$ [3] is actually only true for a bulk superconductor. Recent studies suggests this value should be much smaller for thin films; as low as $\eta \sim 0.4$ for thin-film Al [4], thus we take $\eta = 0.4$.

Any change in the quasiparticle population corresponds to a change in the detector's resonance frequency. Using conventional superconductivity theory we simulate the detector response as a function of change in quasiparticle population $\frac{df}{dN_{qp}}$ to find the maximum resonance frequency shift,

$$df_{max} = N_{qp,xs} \frac{df}{dN_{qp}}. \quad (2)$$

The expected response to single-photon absorption is a pulse with an exponential decay of the resonance frequency shift $df$, The decay time is governed by the dominant time-constant of the detector. We assume this to be the quasiparticle lifetime $\tau_{qp}$, such that

$$df = df_{max} e^{-t/\tau_{qp}}. \quad (3)$$

Our current 1st order model gives an expected pulse height of ~120 Hz.

## 3    Energy Resolution

The energy resolution of a photon detector can be given by

$$\Delta E = h\nu \times \frac{noise}{signal}. \quad (4)$$

The fundamental noise limit of a LEKID is governed by the generation-recombination (GR) noise, in which there is a continuous random fluctuation in the quasiparticle population at any given time: $N_{qp} \pm \sqrt{N_{qp}}$. Therefore, we take the $\sqrt{N_{qp}}$ term as our noise.

When a photon is absorbed, an excess of quasiparticles is created following Eq. (1). We take this to be our signal, such that the fundamental energy resolution limit due to GR noise becomes

**Modelling the Performance of Single-Photon Counting Kinetic Inductance Detectors**

$$\Delta E_{GR} = \frac{\Delta}{\eta}\sqrt{N_{qp}}. \tag{5}$$

Note that this is the energy resolution integrated over the detector time-constant: the quasiparticle lifetime.

Whereas, the fundamental energy resolution for a detector capable of single-photon detection is governed by the Fano limit and given by

$$\Delta E_{photon} = 2\sqrt{2\ln 2}\sqrt{\frac{h\nu F\Delta}{\eta}} = 21 \text{ meV}, \tag{6}$$

where $F$ is the Fano factor [5]. For a LEKID, the Fano factor takes into account the variance in the number of quasiparticles generated due to photon absorption. It is generally accepted to take $F = 0.2$.

It is reasonable to suggest, then, the ultimate energy resolution of a single-photon LEKID is a combination of the two fundamental limits. We therefore propose a maximum energy resolution:

$$\Delta E_{max} = \sqrt{\frac{\Delta^2}{\eta^2}N_{qp} + \left(4 \times 2\ln 2 \times \frac{h\nu F\Delta}{\eta}\right)}. \tag{7}$$

Calculating for our test device, which have meander volumes of ~1400 μm³ and measurements made at 100 mK, we get $\Delta E_{GR} = 0.4$ meV and thus $\Delta E_{max} = 21$ meV. This suggests the detector is at the fundamental photon limit. However, it has been shown there is a limit to the responsivity of a LEKID at low temperatures [6]. The quasiparticle lifetime (and equivalently, the quasiparticle number) saturate at some temperature due to microwave heating [7]. We have measured this saturation in our device, shown in Fig. 2, showing a saturation temperature of approximately 190 mK. This adjusts our energy resolution to $\Delta E_{GR} = 100$ meV and thus from Eq. (7) we get $\Delta E_{max} = 103$ meV. So we are now very firmly in the GR noise limit; which can be improved upon.

## 4  Experimental Set-up and Measurement

We used a LEKID array developed for part of the SpaceKIDs project [8], optimised as a 350 GHz narrowband Earth observation demonstrator. Although not optimised for single-photon detection, it is a well-defined test device with good sensitivity. It is a 624 pixel array, formed of a 30 nm Al film ($T_C = 1.3$ K) on a 320 μm Si substrate. There is an additional TiAl bi-layer on the backside of the substrate which is used as a method to reduce cross talk

**Josie Dzifa Akua Parrianen • Andreas Papageorgiou • Simon Doyle • Enzo Pascale**

[9]. The pixel design follows standard LEKID architecture with the inductive meander patterned into a 3$^{rd}$ order Hilbert fractal, shown in Fig. 1.

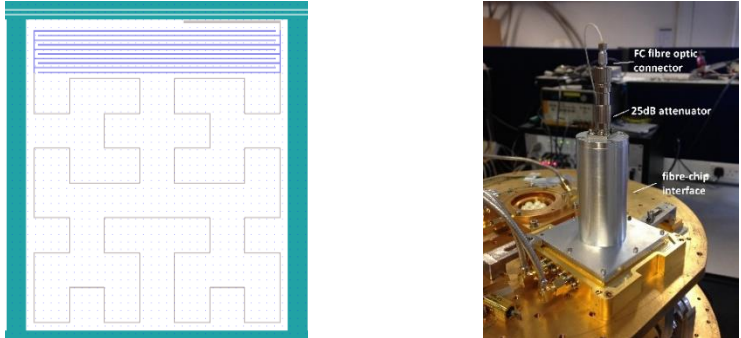

**Fig. 1** *Left* schematic of LEKID architecture; interdigitated capacitor (IDC) finger and meander have linewidth 4 μm with meander volume ~ 1400 μm$^3$. *Right* image of fibre-chip interface. (Colour figure online)

The array was cooled down to a base temperature of 100 mK, in a miniature dilution refrigerator. A standard homodyne readout technique was used to measure the detector response. Measurements were made with a 1550 nm laser diode mounted on the 4 K stage. The incident photons are carried by a 9 μm single-mode fibre optic cable up to the ultra-cold stage and are attenuated before entering the device holder. A modified plate, shown in Fig. 1, ensures roughly half of the pixels are directly illuminated. The device is flood illuminated at a constant DC power level. The change in resonant frequency is measured and compared to when the light source is off.

## 5   Results

The analysis of the data produced in this work has two main aims: i) to measure the impulse response decay time-constant $\tau$ and ii) to identify single-photon absorption events and measuring the pulse amplitude of such events.

The impulse response time-constant was measured by illuminating the detector with square-wave pulses and fitting the pulse decay with an exponential. Stacking of several pulses was used to increase the signal-to-noise ratio (SNR). This procedure was particularly necessary for the significantly noisier data obtained at higher temperatures – in these cases, high accuracy time-alignment was achieved by cross-correlation. Above 200 mK the SNR became too low, at which point we use the traditional method: measuring the noise roll-off from the power spectral density (PSD) of the

**Modelling the Performance of Single-Photon Counting Kinetic Inductance Detectors**

detector response. Fig. 2 shows the combined detector time-constant fitting results, which we assume is dominated by the quasiparticle lifetime.

Another approach to measuring τ was in reconstructing the detector impulse response to single-photon absorption events. Potential absorption events were identified by a step-function match filter. Such events were stacked and averaged, as seen in Fig. 2, which has been fitted with an exponential of τ = 1.9 ms. This matches the same value found for square-wave measurements made at 100 mK.

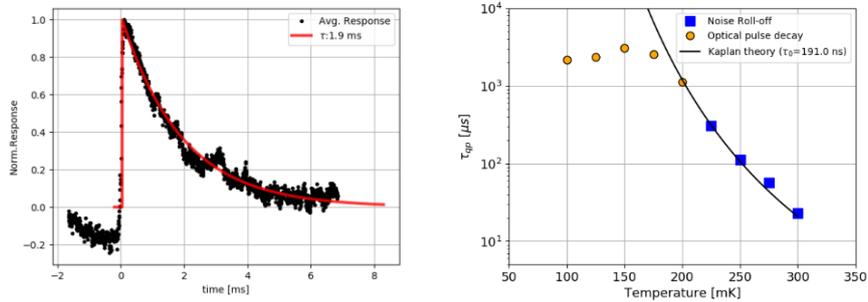

**Fig. 2** *Left* averaged normalised stacked impulse response found in the illuminated detector response. An exponential (red line) is fitted with τ = 1.9 ms. *Right* the quasiparticle lifetime as a function of bath temperature, using two methods for extracting the time-constant. Kaplan theory [10] is fitted to the noise roll-off data (blue). (Colour figure online)

Having obtained an estimate of the impulse response, detector time-streams were Wiener filtered – with bandwidth ~2 kHz – and potential photon absorption events were identified by match filter. The same procedure was applied to both illuminated and dark detector data; the extraction counts are shown in Fig. 3. The illuminated detector detections significantly outnumber the "false detections" – limitations of the impulse extraction procedure – of the dark detector, for impulses of amplitude >200 Hz. This surplus of detections is attributed to true photon absorption events. Given the current extraction method employed here, photon absorption events with amplitude <200 Hz are more difficult to distinguish between true and false detections. This uncertainty is depicted by the vertical error bars shown in Fig. 3.

The expected photon absorption distribution is then measured by subtracting the dark data detections (comprised of false detections only) from the detections of the illuminated data (false detections + photon absorption events), resulting in the plot seen in Fig. 3. Notice the photon absorption distribution peaks at an amplitude of ~120 Hz; matching our expected pulse height. There may also be a peak at ~240 Hz with very small error bars, suggestive of 2-photon absorption events.

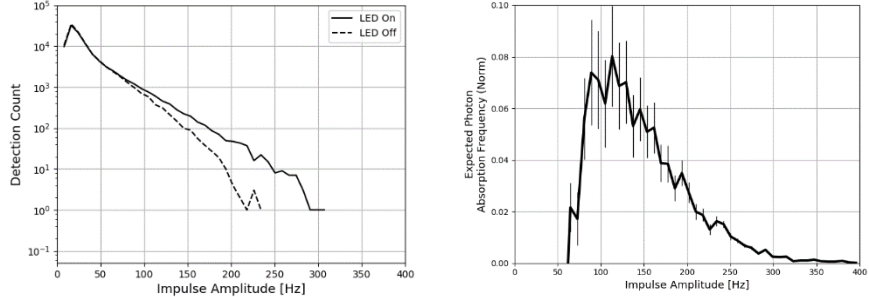

**Fig. 3** *Left* impulse detection counts as a function of impulse amplitude; detection counts shown for both dark (LED off) and illuminated (LED on) detector. Dark detections are false detections and represent the sensitivity limits of the detection algorithm. *Right* normalised expected photon absorption event distribution; which shows a peak at approximately 120 Hz.

As an initial estimate for the energy resolution of our measurements, we refer to Eq. (4); taking the full-width half-maximum (FWHM) as our noise and the peak value as our signal, such that $\Delta E_{calc} = 660$ meV. This is ~6.5 times larger than expected. However, the Weiner filtering process means we only integrate over ~$\tau_{qp}/4$ leading to a $\sqrt{4}$ increase in GR noise contributions: $\Delta E_{GR} = 200$ meV. This brings our estimated energy resolution to ~3 times larger than expected. Also note the FWHM is likely skewed by the apparent ~240 Hz peak and it is probable the distribution includes responses from photons absorbed in the capacitor; causing partial responses when only part of the quasiparticle diffusion occurs in the detecting element: the inductive meander. Therefore we consider this calculation to be an overestimate.

## 6  Conclusions and Future Work

We show confirmation of single-photon detection which corroborates our 1[st] order model as well as the use of $\eta = 0.4$ for thin-film Al. Our detection model will be developed further to include the response of photon absorption in non-detecting elements of the device. We will also continue this work using higher energy photons – to eliminate the false detections – and develop optimised devices as a means of further investigating the mechanisms that enable energy-resolving detection in LEKIDs.

# Modelling the Performance of Single-Photon Counting Kinetic Inductance Detectors

**Acknowledgements** We acknowledge the Science and Technology Facilities Council (STFC) Consolidated Grant Ref: ST/N000706/1 and studentship funding for supporting this work.